\documentclass[twocolumn,showpacs,preprintnumbers,amsmath,amssymb]{revtex4}

\usepackage{graphicx}
\usepackage{dcolumn}
\usepackage{bm}

\begin{document}

\preprint{}

\title{Observation of the Borromean three-body F\"orster resonances \\ for three interacting Rb Rydberg atoms}

\author{D.~B.~Tretyakov$^{1, 2}$}
\author{I.~I.~Beterov$^{1, 2}$}
\author{E.~A.~Yakshina$^{1, 2}$}
\author{V.~M.~Entin$^{1, 2}$}
\author{I.~I.~Ryabtsev$^{1, 2}$}
  \email{ryabtsev@isp.nsc.ru}
\author{P.~Cheinet$^{3}$}
\author{P.~Pillet$^{3}$}

\affiliation{$^1$Rzhanov Institute of Semiconductor Physics SB RAS, 630090 Novosibirsk, Russia }
\affiliation{$^2$Novosibirsk State University, 630090 Novosibirsk, Russia}
\affiliation{$^3$Laboratoire Aime Cotton, CNRS, Univ. Paris-Sud, ENS Paris-Saclay, 91405 Orsay, France}
\date{27 October 2017}

\begin{abstract}

Three-body F\"orster resonances at long-range interactions of Rydberg atoms were first predicted and observed in Cs Rydberg atoms by Faoro et al. [Nature Commun. \textbf{6}, 8173 (2015)]. In these resonances, one of the atoms carries away an energy excess preventing the two-body resonance, leading thus to a Borromean type of F\"orster energy transfer. But they were in fact observed as the average signal for the large number of atoms $N\gg1$. In this Letter we report on the first experimental observation of the three-body F\"orster resonances ${\rm 3}\times nP_{3/2} (|{\rm M}|)\to nS_{1/2} +(n+1)S_{1/2} +nP_{3/2} (|{\rm M}^{*} |)$ in a few Rb Rydberg atoms with $n=36, 37$. We have found here clear evidence that there is no signature of the three-body F\"orster resonance for exactly two interacting Rydberg atoms, while it is present for \textit{N}=3$-$5 atoms. This demonstrates the assumption that three-body resonances can generalize to any Rydberg atom. As such resonance represents an effective three-body operator, it can be used to directly control the three-body interactions in quantum simulations and quantum information processing with Rydberg atoms.

\end{abstract}

\pacs{32.80.Ee, 32.70.Jz , 32.80.Rm, 03.67.Lx}
 \maketitle

Highly excited Rydberg atoms exhibit strong long-range interactions due to their huge dipole moments that grow as $n^2$ with increasing the principal quantum number \textit{n}~[1]. This is especially attractive for the development of quantum computers and simulators based on qubits represented by single alkali-metal atoms in arrays of optical dipole traps or optical lattices [2-5]. In particular, Rydberg-atom-based quantum simulators can directly model various objects in solid-state physics due to their ability to mimic various possible interactions between their constituents, if such interactions in a quantum simulator are appropriately controlled [6-14]. 

Interactions between Rydberg atoms are flexibly controlled by the dc or radio-frequency (rf) electric field via Stark-tuned [15], microwave [16-19], or rf-assisted [4,16] F\"orster resonances corresponding to the F\"orster resonant energy transfer (FRET). F\"orster resonances have been demonstrated to be efficient tools in cold Rydberg atoms [20, 21] to tune interactions in strength and distance and can be either resonant dipole-dipole or nonresonant van der Waals interactions. The interactions are typically described by a two-body operator of dipole-dipole interaction for each pair of atoms in the ensemble [1]. After such an interaction the two atoms are found in an entangled state, so that a measurement over one atom deterministically predicts the state of the other atom. This entanglement is the quantum resource, which is used in quantum computations and simulations [2-13,22].

Some exotic quantum simulations demand to simultaneously control the interactions of three atoms [22-28]. This demands a three-body quantum operator that changes the states of the three qubits simultaneously and makes them all entangled. Three-body operators are described by a combination of two-body operators, which in fact are reduced to a single effective three-body operator. 

Such an operator has been proposed and implemented recently as a Borromean three-body FRET in a frozen Rydberg gas of Cs atoms [29]. In these three-body resonances, one of the atoms carries away an energy excess preventing the two-body resonance, leading thus to a Borromean type of F\"orster energy transfer. Here the Borromean transfer is featured by the strong isolated three-body energy transfer with a negligible contribution of the two-body effect. This allows us to characterize the three-body effect while, it is usually impossible in other systems because it is imbedded in the strong two-body effect signal. The experiment in Ref.~[29] was done with an ensemble of $\sim10^5$ Cs atoms in the interaction volume of $\sim200$~$\mu$m in size. Therefore, the three-body F\"orster resonance was in fact observed as the average signal for the large number of atoms $N\gg1$.

In this Letter, we present the first experimental observation of the Borromean three-body F\"orster resonance ${\rm 3}\times nP_{3/2} (|{\rm M}|)\to nS_{1/2} +(n+1)S_{1/2} +nP_{3/2} (|{\rm M}^{*} |)$ for \textit{N}=3$-$5 Rb Rydberg atoms with $n=36, 37$. We have found clear evidence that there is no signature of the three-body F\"orster resonances for exactly two interacting Rydberg atoms, while it is present for the larger number of atoms. We thus demonstrate the possible generalization of this effect to other Rydberg atoms.

The experiments are performed with cold $^{85}$Rb atoms in a magneto-optical trap [4,30]. Our experiments feature atom-number-resolved measurement of the signals obtained from \textit{N}=1$-$5 detected Rydberg atoms with a detection efficiency of $T\approx\,$70\% [31]. It is based on a selective field ionization (SFI) detector with a channel electron multiplier (CEM) and postselection technique [32]. The electric field for SFI is formed by two stainless-steel plates that are 1~cm apart. These plates have holes covered by meshes for passing the vertical cooling laser beams and the electrons to be detected. The dc electric field, which is homogeneous due to the meshes, is calibrated with 0.2\% uncertainty using the Stark spectroscopy of the microwave transition $37P_{3/2}\to 37S_{1/2}$ at 80.124~GHz [30].

The CEM output pulses from the \textit{nS} and [\textit{nP}+(\textit{n}+1)\textit{S}] states (the two latter states have nearly identical ionizing fields) are detected with two independent gates and postselected over the number of the detected Rydberg atoms \textit{N}=1$-$5. The normalized \textit{N}-atom signals \textit{S$_N$} are the fractions of atoms that have undergone a transition to the final \textit{nS} state.  

In this experiment, the detection of \textit{N} Rydberg atoms means that there were \textit{N} interacting Rydberg atoms with $T\approx $70\% confidence and \textit{N}+1 interacting atoms with $(1-T)\approx $30\% confidence [31]. Therefoere, the recorded F\"orster resonance spectra were additionally processed to extract the true multiatom spectra $\rho_i$ taking into account finite detection efficiency [33]. As shown in our paper [32], for the nonideal SFI detector, which detects fewer atoms than actually have interacted, various true multiatom spectra $\rho_i$ of the F\"orster resonances for $i$ interacting Rydberg atoms contribute to our measured signals $S_N$ for $N$ detected Rydberg atoms to a degree that depends on the mean numbers of the excited and detected atoms. The signals $S_N$ are thus a mixture of the spectra $\rho_i$ from the larger numbers of actually interacted atoms $i\geq N$. In order to derive $\rho_i$ from $S_N$, we have developed a procedure that solves the system of linear equations and approximately expresses each $\rho_i$ via various $S_N$ [33].

The excitation of Rb atoms to the \textit{nP}$_{3/2}$ Rydberg states is realized via the three-photon transition $5S_{1/2} \to 5P_{3/2} \to 6S_{1/2} \to nP_{3/2} $  by means of three cw lasers modulated to form 2 $\mu$s exciting pulses at a repetition rate of 5~kHz [4,35]. A small Rydberg excitation volume of $\sim15$ $\mu$m in size is formed using the crossed tightly-focused laser beams. The laser intensities are adjusted to obtain about one Rydberg atom excited per laser pulse on average. We use a Stark-switching technique [35,36] to switch the Rydberg interactions on and off. Laser excitation occurs during 2 $\mu $s at a fixed electric field of 5.6 V/cm. Then the field decreases to a lower value near the resonant electric field, which acts for 3~$\mu$s until the field increases back to 5.6 V/cm. Then, 0.5 $\mu$s later, a ramp of the strong field-ionizing electric pulse of 200 V/cm is applied. The lower electric field is slowly scanned across the F\"orster resonance and the SFI signals are accumulated for $10^3-10^4$ laser pulses. 

Figure~1 presents the numerically calculated Stark structure of the F\"orster resonance ${\rm 3}\times 37P_{3/2} \to 37S_{1/2} +38S_{1/2} +37P_{3/2}^{*} $ for three Rb Rydberg atoms. The energies W of various three-body collective states are shown versus the controlling dc electric field. The intersections between collective states (labeled by numbers) correspond to the F\"orster resonances of various kinds. Actually, there are the anticrossings at the intersection points due to Rydberg interactions [29]. In our experiment, however, the average two-body dipole-dipole interaction energy is small ($\sim 0.25$~MHz [35]) and the anticrossings are not visible in the energy scale of Fig.~1. 

\begin{figure}
\includegraphics[scale=1]{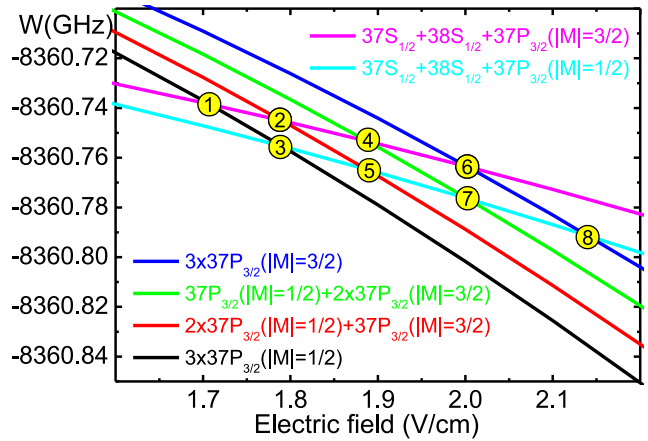}
\caption{\label{Fig1} Numerically calculated Stark structure of the F\"orster resonance ${\rm 3}\times 37P_{3/2} \to 37S_{1/2} +38S_{1/2} +37P_{3/2}^{*} $ for three Rb Rydberg atoms. The energies W of various three-body collective states are shown versus the controlling electric field. Intersections between collective states (labeled by numbers) correspond to the F\"orster resonances of various kinds. Intersections 2-7 are in fact two-body resonances that do not require the third atom. The intersections 1 and 8 are three-body resonances occurring only in the presence of the third atom that carries away an energy excess preventing the two-body resonance. }
\end{figure}

Intersections 2$-$7 are, in fact, two-body resonances that do not require the third atom and can be observed for two atoms. In such resonances, the dipole-dipole interaction induces transitions from the initial 37\textit{P}$_{3/2}$ state to the final 37\textit{S}$_{1/2}$ and 38\textit{S}$_{1/2}$ states in two of the three atoms, while the third atom remains in its initial \textit{P} state that does not change.

Intersections 1 and 8 are three-body resonances occurring only in the presence of the third atom that carries away an energy excess preventing the two-body resonance, leading thus to a Borromean type of F\"orster energy transfer [29]. The three-body resonances are distinguished from the two-body ones by the fact that the third atom does not remain in its initial \textit{P} state as its initial moment projection (\textbar M\textbar  =1/2 or \textbar M \textbar  =3/2) changes to the other one (\textbar M$^*$\textbar  =3/2 or \textbar M$^*$\textbar  =1/2, correspondingly). Therefore, the three-body resonance corresponds to the transition when the three interacting atoms change their states simultaneously. 

In our experiments, cold Rb atoms are excited in the dc electric field either to the initial 37\textit{P}$_{3/2}$(\textbar M\textbar  =1/2) Stark sublevel or to the 37\textit{P}$_{3/2}$(\textbar M\textbar  =3/2) one. Therefore, not all resonances 1$-$8 in Fig.~1 can be observed simultaneously. For the initial state 37\textit{P}$_{3/2}$(\textbar M\textbar =1/2) only  resonances 1 and 3 are observable, while for the initial state 37\textit{P}$_{3/2}$(\textbar M\textbar  =3/2) we can observe only resonances 6 and 8. The intermediate resonances 2, 4, 5 and 7 are observable only when both \textbar M\textbar  =1/2 and \textbar M\textbar  =3/2 atoms are initially excited, as in our earlier paper [37] where we used the excitation by broadband pulsed lasers.

Figures 2(a) and 2(b) show the Stark-tuned F\"orster resonances observed for various numbers of the interacting atoms \textit{i}=2$-$5. In Fig.~2(a) the atoms are in the initial state 37\textit{P}$_{3/2}$(\textbar M\textbar =1/2). The main peak at 1.79 V/cm is the ordinary two-body resonance that occurs for all \textit{i}=2$-$5 and corresponds to intersection 3 in Fig.~1. This resonance was studied in detail in our previous papers [31,35]. The additional peak at 1.71~V/cm is the predicted three-body resonance 1 of Fig.~1 that is absent for \textit{i}=2 and appears only for \textit{i}=3$-$5. The two-body and three-body peak positions well agree with those predicted by Fig.~1.

\begin{figure}
\includegraphics[scale=1]{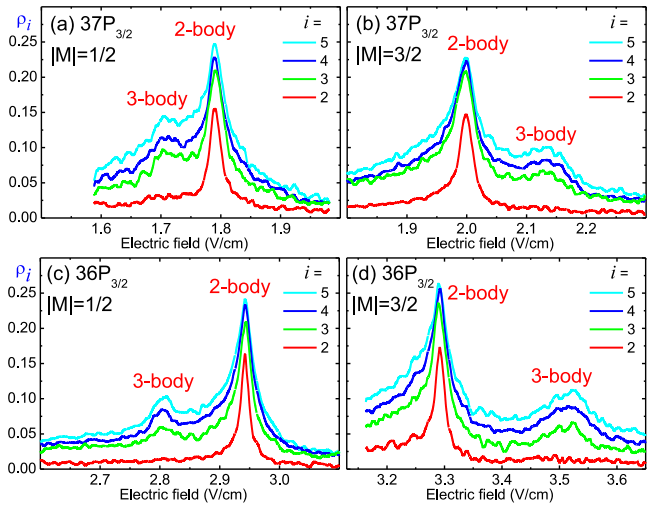}
\caption{\label{Fig2} Stark-tuned F\"orster resonances in Rb Rydberg atoms observed for various numbers of atoms \textit{i}=2$-$5 and various initial states: (a) 37\textit{P}$_{3/2}$(\textbar M\textbar =1/2); (b) 37\textit{P}$_{3/2}$(\textbar M\textbar =3/2); (c) 36\textit{P}$_{3/2}$(\textbar M\textbar =1/2); (d) 36\textit{P}$_{3/2}$(\textbar M\textbar =3/2). The main peaks are two-body resonances, and the additional peaks are three-body resonances. The three-body resonance is absent for \textit{i}=2 in all records, evidencing its three-body nature.}
\end{figure}

The feature at 1.71 V/cm could in principle be caused by the imperfection of the electric-field pulses used to control the F\"orster resonance, as it was observed and discussed in our paper [35]. In order to check for this effect, the resonance has also been recorded for atoms in the initial state 37\textit{P}$_{3/2}$(\textbar M\textbar =3/2), as shown in Fig.~1(b). We see that the three-body resonance changes its position with respect to the two-body resonance, in full agreement with Fig.~1. Again, the main peak at 2.0 V/cm is the ordinary two-body resonance that occurs for all \textit{i}=2$-$5. The additional peak at 2.14 V/cm is the three-body resonance that is absent for \textit{i}=2 and appears only for \textit{i}=3$-$5. We conclude that the three-body resonances really take place, as their positions and behavior well agree with theoretical predictions. The imperfection of the electric-field pulses results only in the slight asymmetry of the main two-body resonances.

Figures 2(a) and 2(b) show that the two-body and three-body resonances partially overlap. This overlapping increases as \textit{i} grows due to the increase of the total interaction energy and broadening of the two-body resonance. The overlapping can be reduced if a lower Rydberg state is used [29]. For example, if we take atoms in the initial state 36\textit{P}$_{3/2}$, the Stark structure of the F\"orster resonance is the same as in Fig.~1, but the separation between intersections 1 and 3 is 140 mV/cm instead of 80 mV/cm for the 37\textit{P}$_{3/2}$ atoms. Figures 2(c) and 2(d) present the two-body and three-body resonances recorded for atoms in the initial state 36\textit{P}$_{3/2}$. The resonances are similar to those in Figs.~2(a) and 2(b), but are better visible due to the larger separation. They additionally confirm that the three-body resonances really take place and can be observed separately from the two-body ones.

In our previous experiments [31,35], we used only atoms in the initial state 37\textit{P}$_{3/2}$(\textbar M\textbar =1/2). Therefore, in the related theoretical analysis [31,35,38] we considered only the two-body resonance 3 of Fig.~1 and ignored the possibility of the three-body resonance 1. As a result, the numerically calculated multiatom spectra $\rho_i$ for \textit{i}=2$-$5 were not disturbed by the three-body resonance and had symmetric line shapes. Our present experiment has revealed that the three-body resonance affects the line shapes for \textit{i}=3$-$5 and causes the asymmetry. This asymmetry indicates that some atoms undergo a nonresonant transition from the initial state 37\textit{P}$_{3/2}$(\textbar M\textbar =1/2) to another Stark sublevel 37\textit{P}$_{3/2}$(\textbar M$^*$\textbar =3/2), but such a transition is not described by the two-body operator of dipole-dipole interaction. This requires a new theoretical model to be developed. It is a rather complicated problem, since we should take into account all Stark and magnetic sublevels of the interacting Rydberg atoms. In this Letter we limited our theoretical considerations only by the cases of two and three interacting Rydberg atoms.

For two Rydberg atoms in the initial state 37\textit{P}$_{3/2}$(\textbar M\textbar =1/2) only one F\"orster resonance 3 of Fig.~1 is possible, which corresponds to the resonant transition between two collective states 2$\times$37\textit{P}$_{3/2}$(\textbar M\textbar =1/2)$\rightarrow$37\textit{S}$_{1/2}$+ 38\textit{S}$_{1/2}$. Its dipole-dipole matrix element is given by  

\begin{equation} \label{Eq1} 
V=\frac{d_{1} d_{2} }{4\pi \varepsilon _{0} } \left[\frac{1}{R^{3} } -\frac{3\, \, Z^{2} }{R^{5} } \right], 
\end{equation}

\noindent where $d_{1}$ and $d_{2} $ are the \textit{z} components of the matrix elements of dipole moments of transitions ${\left| 37P_{3/2} \left({\rm M}=1/2\right) \right\rangle} \to {\left| 37S_{1/2} \left({\rm M}=1/2\right) \right\rangle} $ and ${\left| 37P_{3/2} \left({\rm M}=1/2\right) \right\rangle} \to {\left| 38S_{1/2} \left({\rm M}=1/2\right) \right\rangle} $, $Z$ is the \textit{z }component of the vector \textbf{R} connecting the two atoms (\textit{z} axis is chosen along the dc electric field), and $\varepsilon_0$ is the dielectric constant. For the weak interaction, the two-body F\"orster resonance amplitude is $\rho _{2} \sim V^{2} $ [35].

For three Rydberg atoms in the initial state 37\textit{P}$_{3/2}$(\textbar M\textbar =1/2), the two F\"orster resonances 1 and 3 of Fig.~1 are possible. The three-body resonance 1 corresponds to the resonant transition between collective states 3$\times$37\textit{P}$_{3/2}$(\textbar M\textbar =1/2)$\rightarrow$37\textit{S}$_{1/2}$+ 38\textit{S}$_{1/2}$+37\textit{P}$_{3/2}$(\textbar M$^*$\textbar =3/2). This transition is, in fact, composed of the two nonresonant two-body relay transitions 3$\times$37\textit{P}$_{3/2}$(\textbar M\textbar =1/2)$\rightarrow$37\textit{S}$_{1/2}$+ 38\textit{S}$_{1/2}$+37\textit{P}$_{3/2}$(\textbar M\textbar =1/2)$\rightarrow$37\textit{S}$_{1/2}$+38\textit{S}$_{1/2}$+ 37\textit{P}$_{3/2}$(\textbar M$^*$\textbar =3/2) occurring simultaneously. The latter occurs due to non-resonant exchange interaction \textit{nP}$_{3/2}$(M)\textit{+n$'$S}$\rightarrow$\textit{ n$'$S+nP}$_{3/2}$(M$^*$) corresponding to the excitation hopping between \textit{S} and \textit{P} Rydberg atoms [29,38]. Despite the use of a relay, the transfer occurs in a single step, implying a Borromean character of the relay atom which absorbs the energy of the finite F\"orster defect. The perturbation theory shows that for the weak interaction the three-body F\"orster resonance amplitude is  $\rho _{3} \sim (VV^{*} /\Delta )^{2} $, where $V^{*} $ is the same as \textit{V} but for the transitions ${\left| 37P_{3/2} \left({\rm M}=3/2\right) \right\rangle} \to {\left| 37S_{1/2} \left({\rm M}=1/2\right) \right\rangle} $ and ${\left| 37P_{3/2} \left({\rm M}=3/2\right) \right\rangle} \to {\left| 38S_{1/2} \left({\rm M}=1/2\right) \right\rangle} $, and $\Delta$/(2$\pi$)=9.5 MHz is the energy splitting between 37\textit{P}$_{3/2}$(\textbar M\textbar =1/2) and 37\textit{P}$_{3/2}$(\textbar M\textbar =3/2) Stark sublevels in the electric field of 1.71 V/cm. 

\begin{figure}
\includegraphics[scale=1]{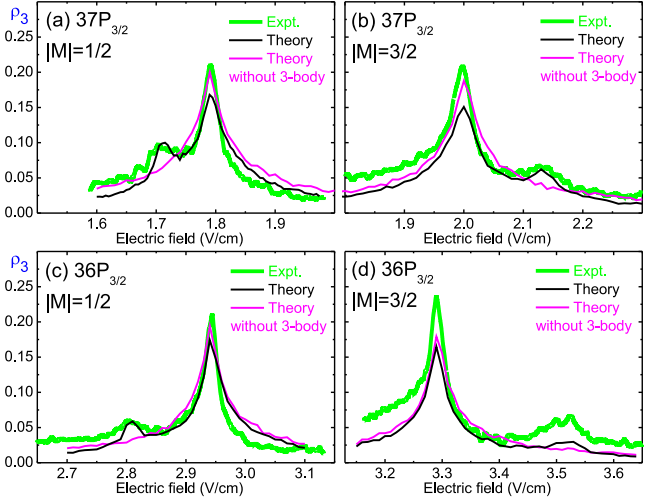}
\caption{\label{Fig3} Comparison between the theory and experiment for the three-atom Stark-tuned F\"orster resonances ${\rm 3}\times nP_{3/2} (|{\rm M}|)\to nS_{1/2} +(n+1)S_{1/2} +nP_{3/2} (|{\rm M}^{*} |)$ in Rb Rydberg atoms for the initial states: (a) 37\textit{P}$_{3/2}$(\textbar M\textbar =1/2); (b) 37\textit{P}$_{3/2}$(\textbar M\textbar =3/2); (c) 36\textit{P}$_{3/2}$(\textbar M\textbar =1/2); (d) 36\textit{P}$_{3/2}$(\textbar M\textbar =3/2). The theoretical spectra have been calculated for the cubic interaction volume of 15$\times$15$\times$15 $\mu$m$^3$, 3 $\mu$s interaction time and Monte Carlo averaging over 1000 random atom positions. The thick green (gray) lines are the experiment, the thin black lines are the full theory, and the thin magenta (dark gray) lines are the theory without accounting for the three-body resonances.}
\end{figure}

The three-body resonance is thus less effective than the two-body one at the weak dipole-dipole interaction ($V<\Delta $). However, when the three-body resonance is exactly tuned, its contribution to the population transfer generally exceeds the contribution from the two-body interaction, which is offresonant in this case. The condition for the three-body resonance to be of the Borromean type is thus satisfied.

We have done numerical simulations of the experimental F\"orster resonances of Fig.~2 for \textit{i}=3 atoms using the method described in Refs.~[31,38]. It is based upon solving the Schr\"odinger's equation with subsequent Monte Carlo averaging over the random positions of the three atoms in a single interaction volume. The Stark and Zeeman structures of all Rydberg states are fully taken into account. The numerical results and their comparison with the experimental data of Fig.~2 are presented in Fig.~3. The thick green (gray) lines are the experimental three-atom data, the thin black lines are the full theory, and the thin magenta (dark gray) lines are the theory without accounting for the three-body resonances. The theoretical spectra have been calculated and averaged over 1000 random atom positions for the cubic interaction volume of 15$\times$15$\times$15~$\mu$m$^3$ and 3 $\mu$s interaction time, which correspond to our experimental parameters. 

The overall agreement of the full theory with the experiment in Fig.~3 is satisfactory. The calculated line shapes of the two-body resonances are close to the experimental ones. These are cusp-shaped resonances that are formed upon spatial averaging in a single interaction volume, as discussed in our paper [35] and other papers [39,40]. When the three-body resonances are not accounted for by the theory, the height of the two-body peak grows because the population does not leak to the other three-atom states, while the three-body peaks are absent at all. The three-body resonances are well reproduced by theory in Figs.~3(a)-3(c), in both their heights and widths. 

However, some discrepancy between the experiment and theory is found for the 36\textit{P}$_{3/2}$(\textbar M\textbar =3/2) state atoms in Fig.~3(d). This case is distinguished by the largest separation $\Delta$ between the two-body and three-body resonances. The theory predicts weaker two- and three-body F\"orster resonances than those observed experimentally. One of the explanations could be that the Schr\"odinger equation model gives incorrect time dynamics of the populations at large $\Delta$. This discrepancy points towards the need to build a new model based on the density-matrix equations, as we did for \textit{i}=2 atoms in Ref.~[35]. Compared to the Schr\"odinger equation, the density-matrix model gives a faster time dynamics of the populations in the presence of additional dephasing (unresolved hyperfine structure of Rydberg states and fluctuations of the controlling electric field as observed in Ref.~[35]). But building this model is a complicated task which requires a dedicated study because of the huge number of collective states if the Stark and Zeeman structures are accounted for. 

In conclusion, our experiments with a few Rb Rydberg atoms in various initial states have clearly shown the need for three atoms to obtain a three-body resonance signature in perfect agreement with expectations. The three-body resonance corresponds to a transition when the three interacting atoms change their states simultaneously (two atoms go to the \textit{S} states, and the third one remains in the \textit{P} state but changes its moment projection). Such a Borromean-type transfer displays strong three-body energy transfer with a negligible contribution of two-body transfer. As the three-body resonance appears at the different dc electric field with respect to the two-body resonance, it represents an effective three-body operator, which can be used to directly control the three-body interactions. This can be especially useful in quantum simulations and quantum information processing with neutral atoms in optical lattices [2-15]. It can also allow us to test and study a quantum system where the basic interaction is a three-body interaction.

We note that the Borromean trimers of Rydberg atoms have been predicted in Ref.~[41], and excitation transfer in a spin chain of three Rydberg atoms has been observed experimentally in Ref.~[42]. We also note that, in principle, it is possible to organize three-body interactions for almost arbitrary Rydberg states using the radio-frequency-assisted F\"orster resonances occurring between Floquet sidebands of Rydberg states in a radiofrequency electric field [43]. Finally, F\"orster resonances of the higher orders (four-body etc.) can also be observed in the electric field which is different from the two-body one [29,34].

The authors are grateful to Elena Kuznetsova and Mark Saffman for fruitful discussions. This work was supported by the RFBR Grants No. 16-02-00383 and No. 17-02-00987, the Russian Science Foundation Grant No. 16-12-00028 (for laser excitation of Rydberg states), the Siberian Branch of RAS, the Novosibirsk State University, the public Grant CYRAQS from Labex PALM (ANR-10-LABX-0039) and the EU H2020 FET Proactive project RySQ (Grant No.~640378).

\appendix*

\section{SUPPLEMENTARY MATERIAL} 

\subsection*{Derivation of the true many-body spectra from the experimental multiatom spectra of the F\"orster resonances}

The measured normalized \textit{N}-atom signals $S_{N} $ are the fractions of atoms that have undergone a transition to the final \textit{nS} state (or the population of the \textit{nS} state per atom). As shown in our paper [S1], for the nonideal selective-field-ionization (SFI) detector, which detects fewer atoms than actually have interacted, various true multiatom spectra $\rho _{i} $ of the F\"orster resonances for \textit{i} interacting Rydberg atoms contribute to our measured signals $S_{N} $ for \textit{N} detected Rydberg atoms to a degree that depends on the mean number of the detected atoms. The signals $S_{N} $ are thus a mixture of the spectra $\rho _{i} $ from the larger numbers of actually interacted atoms $i\ge N$:

\begin{equation} \label{EqA1} 
S_{N} =\, \rho +e^{-\bar{n}(1-T)} \sum _{i=N}^{\infty }\rho _{i} \frac{\left[\bar{n}(1-T)\right]^{i-N} }{(i-N)!}  ,      
\end{equation} 

\noindent where $\rho $ is a nonresonant background signal due to blackbody-radiation-induced transitions and background collisions, $\bar{n}$ is the mean number of Rydberg atoms excited per laser pulse, and \textit{T} is the detection efficiency of the SFI detector. The value of $\rho $ should be the same for various \textit{N} since it is caused by the parasitic transitions in each single atom.

The mean number of detected Rydberg atoms is $\bar{n}T$. The measurement of this value and of the relationship 

\begin{equation} \label{EqA2} 
\alpha =(S_{1} -\rho )/(S_{2} -\rho ) 
\end{equation} 

\noindent at zero F\"orster detuning can provide a measurement of the unknown values of $\bar{n}$ and \textit{T}. In Ref.~[S1] we considered the case of the weak dipole-dipole interaction, when the following scaling was assumed to be valid:

\begin{equation} \label{EqA3} 
\rho _{i} \approx (i-1)\, \rho _{2} ,    
\end{equation} 

\noindent For this case it was shown that

\begin{equation} \label{EqA4} 
\bar{n}\approx [\alpha /(1-\alpha )+\bar{n}T].   
\end{equation} 

\noindent This expression, however, is valid only for the very weak dipole-dipole interaction, when multiatom F\"orster resonances are far below the saturation, as it was in our experiment with the Na thermal atomic beam [S1].
 
\begin{figure}
\includegraphics[scale=1]{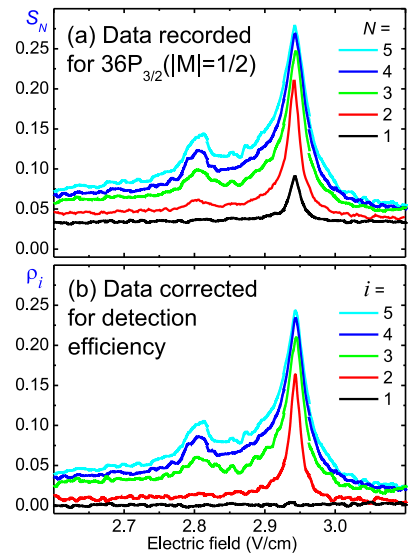}
\caption{\label{Fig4} (color online) (a) Raw data $S_N$ recorded for the Stark-tuned F\"orster resonance in Rb Rydberg atoms for various numbers of the detected atoms \textit{N}=1$-$5 and initial state 36\textit{P}$_{3/2}$(\textbar M\textbar =1/2). Presence of the resonance for \textit{N}=1 is due to the finite detection efficiency of 72\%. (b) True multiatom spectra $\rho_i$ derived from $S_N$. The data are corrected for the detection efficiency using the procedure described in the text.}
\end{figure}

Now let us consider an example of the multiatom F\"orster resonance for $36P_{3/2} (|{\rm M|}=1/2)$ atoms recorded for the interaction time of 3 $\mu$s in our present experiment, shown in Fig.~4(a). Our aim is to make a decomposition of the experimental records $S_1-S_5$ for \textit{N}=1$-$5 detected Rydberg atoms in order to obtain the true multiatom spectra $\rho _{i} $ of the F\"orster resonances for exactly \textit{i} interacting Rydberg atoms, which are defined according to Eq.~\eqref{EqA1}. For this purpose, we first need to find the unknown values of $\bar{n}$ and \textit{T}. 

As a starting point, for this experiment we already know the mean number of the detected Rydberg atoms $\bar{n}T\approx 1.05$, which was specially measured and recorded in each experiment. Then, using Eq.~\eqref{EqA4} we in principle can find $\bar{n}$ and \textit{T}. However, Eq.~\eqref{EqA4} seems to be invalid for Fig.~4(a), because the spectra are close to the saturation and Eq.~\eqref{EqA3} obviously does not work. Therefore we need first to modify Eq.~\eqref{EqA4} for the case of saturation.

Figure 5(a) presents the results of numerical simulations for the theoretical multiatom two-body spectra $\rho _{i} $ for the 36\textit{P}$_{3/2}$(\textbar M\textbar =1/2) atoms in the cubic interaction volume of 17$\times$17$\times$17 $\mu$m$^3$ for the interaction time of 3~$\mu$s (these parameters are close to the experimental ones). It is seen that at zero detuning the amplitudes of all resonances saturate at the 0.25 value. Therefore, at zero detuning instead of Eq.~\eqref{EqA3} we should now adopt that $\rho _{2} \approx \rho _{3} \approx \rho _{4} \approx \rho _{5}\approx ...\, $. Then Eqs. \eqref{EqA1} and \eqref{EqA2} give 

\begin{equation} \label{EqA5} 
\alpha \approx 1-e^{-\bar{n}(1-T)},  
\end{equation} 

\begin{equation} \label{EqA6} 
\bar{n}\approx \ln \frac{1}{1-\alpha } +\bar{n}T. 
\end{equation} 

\begin{figure}
\includegraphics[scale=1]{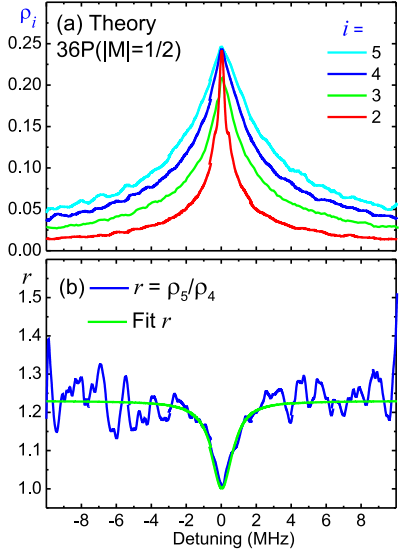}
\caption{\label{Fig5} (color online) (a) Numerical simulation of the two-body F\"orster resonance in Rb Rydberg atoms for various numbers of the interacting atoms \textit{i}=2$-$5 in the initial state 36\textit{P}$_{3/2}$(\textbar M\textbar =1/2). The theoretical spectra have been calculated with the Schr\"odinger's equation for the cubic interaction volume of 17$\times$17$\times$17 $\mu$m$^3$, 3 $\mu$s interaction time and Monte Carlo averaging over 1000 random atom positions. (b) Ratio of the spectra $\rho_5$ and $\rho_4$ is shown by the blue (dark grey) curve, and its fit by the inverted Lorentz function is shown by the green (light grey) curve.}
\end{figure}

The values of $\rho \approx 0.029$, \textit{S}$_1$=0.09 and \textit{S}$_2$=0.21 have been measured from the spectra in Fig.~4(a). This allows us to find $\alpha \approx 0.34$, $\bar{n}\approx 1.46$, and $T\approx 0.72$. With these values we can explicitly write down the expansion coefficients for the multiatom spectra in Eq.~\eqref{EqA1}:

\begin{equation} \label{EqA7} 
\begin{array}{l} {S_{1} =\rho +0.66\rho _{1}+0.27\rho _{2} +0.06\rho _{3} +0.01\rho _{4} +...\,,} \\ {S_{2} =\rho +0.66\rho _{2} +0.27\rho _{3} +0.06\rho _{4} +0.01\rho _{5} +...\,,} \\ {S_{3} =\rho +0.66\rho _{3} +0.27\rho _{4} +0.06\rho _{5} +0.01\rho _{6} +...\,,} \\ {S_{4} =\rho +0.66\rho _{4} +0.27\rho _{5} +0.06\rho _{6} +0.01\rho _{7} +...\,,} \\ {S_{5} =\rho +0.66\rho _{5} +0.27\rho _{6} +0.06\rho _{7} +0.01\rho _{8} +...\,.} \end{array} 
\end{equation} 

\noindent In Eqs.~\eqref{EqA7} we should take into account that $\rho_1=0$ in $S_1$, because there is no interaction for a single atom.

In order to derive $\rho _{2} $ and $\rho _{3} $, which are necessary for the analysis of the three-body F\"orster resonance, we should simplify Eqs.~\eqref{EqA7} to exclude the terms with large numbers of atoms. First, the terms with the weight of 0.01 have small contribution and with a small error can be just added to the preceding terms as follows:

\begin{equation} \label{EqA8} 
\begin{array}{l} {S_{1} =\rho +0.27\rho _{2} +0.07\rho _{3}\,, } \\ {S_{2} =\rho +0.66\rho _{2} +0.27\rho _{3} +0.07\rho _{4}\,, } \\ {S_{3} =\rho +0.66\rho _{3} +0.22\rho _{4} +0.07\rho _{5} \,,} \\ {S_{4} =\rho +0.66\rho _{4} +0.27\rho _{5} +0.07\rho _{6}\,, } \\ {S_{5} =\rho +0.66\rho _{5} +0.27\rho _{6} +0.07\rho _{7}\,. } \end{array} 
\end{equation} 

Second, we believe that the multiatom spectra in Fig.~4(a) are reliably measured for \textit{N}=1$-$4, while the spectrum for \textit{N}=5 can be affected by the nonlinearity of our channeltron. Therefore, in the further analysis we will consider only the experimental spectra with \textit{N}=1$-$4 and need to exclude $\rho _{5} $ and $\rho _{6} $ in Eqs.~\eqref{EqA8}. This can be done if we approximately express $\rho _{5} $ and $\rho _{6} $ via $\rho _{4} $ using the theoretical curves in Fig.~5(a). Figure 5(b) shows as the blue (dark grey) curve the ratio $r=\rho _{5} /\rho _{4} $ taken from Fig.~5(a). This ratio depends on the detuning: it is 1 at zero detuning due to saturation and 1.23 at large detunings. The fluctuations of \textit{r} at large detunings in Fig.~5(b) are due to insufficient statistics of the averaging of small signals, which can be smoothed if the statistics increases or using the fitting function.We have found a fitting function for this dependence [green (grey) curve in Fig.5(b)]:

\begin{equation} \label{EqA9} 
r(\Delta )\approx 1.23-0.23\frac{0.5}{0.5+\Delta ^{2} }\,,  
\end{equation} 

\noindent where detuning $\Delta $ is defined by the electric field \textit{F}(V/cm) for the 36\textit{P}$_{3/2}$(\textbar M\textbar =1/2) atoms as

\begin{equation} \label{EqA10} 
\Delta ({\rm MHz})=-229.73+2.93F+25.494F^{2}\,.  
\end{equation}  

\noindent In the further analysis we take $\rho _{5} \approx \rho _{4} r(\Delta )$. We can also adopt with some precision that $\rho _{6} \approx \rho _{5} r(\Delta )\approx \rho _{4} r^{2} (\Delta )$ in Eqs.~\eqref{EqA8}, although we did not calculate $\rho _{6} $ directly.

With the above assumptions Eqs.~\eqref{EqA8} are modified as 

\begin{equation} \label{EqA11} 
\begin{array}{l} {S_{1} =\rho +0.27\rho _{2} +0.07\rho _{3}\,, } \\ {S_{2} =\rho +0.66\rho _{2} +0.27\rho _{3} +0.07\rho _{4}\,, } \\ {S_{3} =\rho +0.66\rho _{3} +[0.27+0.07r(\Delta )]\rho _{4}\,, } \\ {S_{4} =\rho +[0.66+0.27r(\Delta )+0.07r^{2} (\Delta )]\rho _{4}\,. } \end{array} 
\end{equation} 

The straightforward calculations with Eqs.~\eqref{EqA11} give us the true multi-atom spectra $\rho _{2}-\rho _{4} $ expressed via the measured value of $\rho$ and spectra $S_2-S_4$ of Fig.~4(a):

\begin{equation} \label{EqA12} 
\begin{array}{l} {\rho _{4} \approx \displaystyle \frac{S_{4} -\rho }{0.66+0.27r(\Delta )+0.07r^{2} (\Delta )} }\,, \\ \\ {\rho _{3} \approx \displaystyle \frac{S_{3} -\rho }{0.66} -[0.41+0.1r(\Delta )]\rho _{4} } \,,\\ \\ {\rho _{2} \approx \displaystyle \frac{S_{2} -\rho }{0.66} -0.41\rho _{3} -0.1\rho _{4} }\,. \end{array} 
\end{equation} 

In order to derive Eqs.~\eqref{EqA12} we used only the equations for $S_2-S_4$ in Eqs.~\eqref{EqA11}. But after calculating $\rho _{2} $ and $\rho _{3} $ with Eqs.~\eqref{EqA12} we should also check for the identity

\begin{equation} \label{EqA13} 
\begin{array}{l} {\rho_1 \approx (S_{1} -\rho -0.272\rho _{2} -0.064\rho _{3})/0.66 \approx 0} \end{array},   
\end{equation} 

\noindent which means that we correctly decomposed the measured spectra $S_2-S_4$ into true multiatom spectra.

Figure 5(b) presents the true multiatom spectra $\rho _{i} $ derived from Fig.~5(a) with Eqs.~\eqref{EqA12}. The black curve for $\rho _{1} $ represents the identity of Eq.~\eqref{EqA13}. We see that in the 2-atom spectrum the feature at 1.71 V/cm has disappeared, while in the 3-atom spectrum it is still present. It indicates that in this experiment we really observe the Borromean three-body resonance. The validity of the above considerations is confirmed by the fact that the identity of Eq.~\eqref{EqA13} is well satisfied in Fig.~4(b), being nearly zero. 

The other experimental records (Fig.~2 of the main paper) have been processed in the same way as the records in Fig.~4. We note that the approach we have used here is similar to the approach we applied  earlier to decompose the selective-field-ionization signals from four-body resonances in Cs Rydberg atoms [S2].

\begin{itemize}

\item[\sffamily [S1]] I.~I.~Ryabtsev, D.~B.~Tretyakov, I.~I~.Beterov, and V.~M.~Entin, Effect of finite detection efficiency on the observation of the dipole-dipole interaction of a few Rydberg atoms, Phys. Rev. A \textbf{76}, 012722 (2007); Erratum: Phys. Rev. A \textbf{76}, 049902(E) (2007).

\item[\sffamily [S2]] J.~H.~Gurian, P.~Cheinet, P.~Huillery, A.~Fioretti, J.~Zhao, P.~L.~Gould, D.~Comparat, and P.~Pillet, Observation of a Resonant Four-Body Interaction in Cold Cesium Rydberg Atoms, Phys. Rev. Lett. \textbf{108}, 023005 (2012).

\end{itemize}

\end{document}